
\input jnl.tex
\input reforder.tex
\input eqnorder.tex


\def \gapx {{\hbox {$\  {>\atop \sim}\ $}}}

\def \vopo {(VO)$_2$P$_2$O$_7$\ }
\def \cuno {Cu(NO$_3$)$_2\cdot {5\over 2}$H$_2$O\ }
\def \cunon {Cu(NO$_3$)$_2\cdot {5\over 2}$H$_2$O}
\def \vopon {(VO)$_2$P$_2$O$_7$}


\null
\vskip -.75in
{\singlespace
\smallskip
\rightline{\it ORNL-CCIP-94-05 / RAL-94-044}
\rightline{May 1994}

\title
INELASTIC NEUTRON SCATTERING FROM THE SPIN LADDER COMPOUND \vopo
\vskip 0.2in

\author{R.S.Eccleston,$^1$ T.Barnes,$^2$ J.Brody$^3$ and J.W.Johnson$^3$}
\vskip 0.2in

\affil
$^1$The ISIS Facility, Rutherford Appleton Laboratory,
Chilton, Didcot, OXON OX11 0QX, UK

\affil
$^2$Physics Division and Center for Computationally Intensive Physics,
Oak Ridge National Laboratory, Oak Ridge, TN 37831, USA
and
Department of Physics and Astronomy,
University of Tennessee, Knoxville, TN 37996, USA

\affil
$^3$Corporate Research Laboratories,
Exxon Research and Engineering Company,
Annandale, NJ 08801, USA

\abstract
{
We present results from an inelastic neutron scattering experiment on the
candidate Heisenberg spin ladder vanadyl pyrophosphate, \vopon. We find
evidence for a spin-wave excitation gap of $E_{gap} = 3.7\pm 0.2$ meV, at a
band minimum near $Q=0.8$ \AA$^{-1}$. This is consistent with expectations for
triplet spin waves in \vopo in the spin-ladder model, and is to our knowledge
the first confirmation in nature of a Heisenberg antiferromagnetic spin ladder.
}

\vskip 0.2truecm

\line{PACS Indices:  75.10.Jm, 75.25.+z, 75.30.Ds, 75.50.Ee. \hfill }

\endtitlepage

\doublespace

\vfill
\eject

\centerline{\bf I.Introduction}
\vskip 0.25cm

Many neutron scattering studies of the magnetic excitation spectra
of
Heisenberg spin chains have been reported since Haldane
first suggested that integer- and
half-integer spins might lead to qualitatively different
physics.\refto{Haldane,Affleck}
This conjecture, that only half-integer-spin chains are gapless,
is now generally regarded as experimentally confirmed.

To gain insight into the physics underlying the surprising one-dimensional
result, one may study systems intermediate between spin chains
and two-dimensional systems, which are gapless for all spins. The
simplest of these is the ``Heisenberg
spin ladder", which has isotropic couplings $J_{||}$ along the chains and
$J_{\perp}$ between them. Although this system
with both
antiferromagnetic
\reftorange{3}{4-14}{15}
and
ferromagnetic\reftorange{16}{17-19}{20} interchain
couplings
has been the subject of
considerable recent theoretical interest,
to our knowledge no examples have
previously been confirmed in nature.

Copper nitrate, \cunon, was an early ladder candidate,
since a ladder superexchange
pathway was evident, and the specific heat and
susceptibility could be described by a ladder model. However,
theoretical studies showed that
ladders and alternating chains lead to very similar
thermodynamics,\reftorange{21}{22}{23}
and the issue was finally decided in favor of an
alternating chain by
proton magnetic resonance\refto{cuno1} and elastic neutron
scattering\refto{cuno2}
experiments.
Although
copper hydrazinium chloride\refto{chc1,chc2} is not a complete spin ladder,
it apparently does form a closely related
antiferromagnetic ``sawtooth",
in which Cu$^{+2}$ and Cl$^-$ ions occupy alternate ladder sites.

Recently two other candidate ladder systems have been discussed in the
literature, a Sr$_{n-1}$Cu$_{n+1}$O$_{2n}$ series\refto{SCO1,SCO2}
(expected to consist of coupled ladders with a frustrated
``trellis" coupling\refto{11}), a
La$_{4+4n}$Cu$_{8+2n}$O$_{14+8n}$ series\refto{LCO} (another coupled-ladder
system),
and the isolated-ladder
antiferromagnet vanadyl
pyrophosphate,\reftorange{26}{27-33}{34} \vopon, which is the
subject of this study.

A ladder configuration of
$S=1/2$ V$^{+4}$ ions is clearly evident in
the crystal structure of \vopo\refto{30,32,34}
(see especially Fig.2 of Ref.[\cite{34}]),
with spacings of 3.19(1) \AA \  between rung ions and 3.864(2)\AA \  between
chain ions.
The susceptibility of this material is fitted surprisingly well by an
alternating chain model\refto{34} with $J_1=11.3$ meV and $J_2=7.9$ meV,
but this need not eliminate the ladder model if
the thermodynamic
similarities previously noted for \cuno
apply in the \vopo parameter regime as well.
A recent theoretical study\refto{8}
confirmed this similarity and found that
the \vopo susceptibility was also accurately described by a ladder model with
nearly equal interactions,
$J_{||}=7.76$ meV and
$J_{\perp}=7.82$ meV. The singlet-triplet
gap at $J_\perp/J_{||} = 1.0077$ is given by\refto{SRWunpub}
$E_{gap}/J_{||}  = 0.50759(1)$, corresponding to
$E_{gap}=3.94$ meV with these parameters.

One may test the ladder model of \vopo directly in a neutron scattering
experiment, through a determination of
characteristic propeties of the spin-wave
excitation spectrum.
In the ladder model with the parameters quoted above
the
singlet-triplet gap was predicted\refto{8} to be $\approx 3.9$ meV
at a momentum transfer of
$Qa=\pi$ (equivalent to $Q = 0.813$~{ \AA$^{-1}$}
in \vopon). This one-magnon band
reaches a
maximum energy of 16 meV near
$0.3  $ \AA $^{-1} $ and
$1.3  $ \AA $^{-1}$, and a crossing
spin-triplet two-magnon band gives a secondary gap of
7.9~meV at $Qa=0$ and $2\pi$ and a broad plateau at 17-18 meV
centered on $Qa=\pi$.
(See Fig.6 of Ref.[\cite{8}].)
Structure factor calculations
on finite clusters\refto{JRunpub} indicate that the
strongest neutron scattering should occur near
$Qa=\pi$, and higher levels above these two triplet bands should not
be strongly excited.
In contrast, a fit of the alternating chain
model to the susceptibility predicts a somewhat higher gap of 4.9 meV,
and the associated momentum transfer is problematical (it depends on an
unspecified alternating-chain pathway) and need not have any simple relation to
the $Q=0.813$ \AA$^{-1}$ expected in the ladder model.

\vfill\eject

\centerline{\bf II. Experimental Procedure and Results}
\vskip 0.25cm

At present, single crystals of vanadyl pyrophosphate of a sufficient volume
for neutron inelastic scattering experiments are not available, however
several experiments have demonstrated that neutron scattering
experiments on polycrystalline samples of one-dimensional magnetic
systems can yield considerable information.\refto{MPMSCT}  This is
particularly true when using time-of-flight spectrometers, which can
access a wide range of $(\vec Q,\omega )$ simultaneously.
The powder average
of the correlation function observed experimentally is given by\refto{MPMSCT}

$$
S(Q,\omega) =
\, T(\omega)\, |F(Q^2)|\,
{1 \over 4\pi Q^2}\;
\int_{\vec Q = \vec q_{||} + \vec q_\perp, |\vec Q|=Q}
S(\vec q_{||},\vec q_\perp,\omega)\,
\  d\vec Q
\eqno(1)
$$
where $T(\omega)$ is the temperature factor
$$
T(\omega)=\Big( 1-\exp(- \hbar \omega/k_BT)\Big)^{-1} \ ,
\eqno(2)
$$
$F(Q)$ is
the ionic form factor, and $\vec q_{||}$ and $\vec q_\perp$
are the parallel and perpendicular projections
of the total momentum
transfer $\vec Q$
relative to the intrinsic 1D magnetic axis.
Clearly for any given $Q$ all values of $|\vec q_{||}|\leq Q$ will
contribute to the scattered spectrum. Thus a mode which requires
$|\vec q_{||}|=q_1$
along the 1D axis to
be excited will first appear experimentally at $Q=q_1$, but will persist
at values $Q>q_1$, because these can excite systems with
oblique orientations
such that $|\vec q_{||}|=q_1$.

Measurements of the excitation spectrum of a polycrystalline sample of
\vopo were performed on
HET, a direct geometry chopper spectrometer on the ISIS
pulsed neutron source at the Rutherford Appleton Laboratory.
A rotating Fermi chopper phased to the neutron pulse
monochromates the incoming neutron beam.   Banks of detectors at $4$ m
and $2.5$ m cover the angular ranges of $\phi=3^o$ to $7^o$
and $\phi=9^o$ to $29^o$
respectively.  A further two banks at these lengths are situated at average
angles of $118^o$ and $130^o$, and are used for estimating non-magnetic
background scattering.

The sample was prepared by Brody,
in a manner similar to that described by Johnston,
Johnson, Goshorn and Jacobson.\refto{34} The final annealing was performed
under flowing He at 973 K for 7 days.
For this experiment it was mounted on the cold finger of a closed
cycle refrigerator in a sachet of aluminium foil.
All data described hereafter
were collected while the sample was at a temperature of 13 K,
using incident energies of 100 meV, 35 meV, 25 meV and 15 meV.

Data collected simultaneously across the full angular range of the forward
scattering banks using an incident energy of 25 meV
were combined to produce the contour plot of the
scattering intensity shown in Fig.1. The dispersion relation predicted
by Barnes and Riera\refto{8}
has been superimposed on the plot to
compare the ladder model to the
experimental data, and to demonstrate the effects of powder averaging,
which spreads the scattering over larger $Q$ values than the minimum required
for excitation.
The predicted cut-off of the scattering at approximately 18 meV energy
transfer is
clearly seen, and at lower energies
there is evidence of scattering consistent with the
theoretical dispersion relation, particularly in the region
$Q\gapx 0.8$ \AA$^{-1}$;
the one-magnon band can be seen as it passes through a
minimum energy near $Q=0.8$ \AA$^{-1}$,
where there is an increase in intensity as would be
expected from both a peak in the density of states and in $S(\vec Q,\omega)$.

Using a
lower incident energy of 15 meV the resolution is sufficient to resolve a
gap between the elastic peak and the inelastic
magnetic scattering.  Fig.2 shows
data summed over an angular range of $\phi=14.3^o$ to $18.7^o$,
which
at an energy transfer of 3.7 meV
represents a $Q$ range of
$0.72$ \AA$^{-1}$
to
$0.88$ \AA$^{-1}$ (see Fig.2 insert).
The data have been corrected for $k_f/k_i$ and detector efficiency.
The figure clearly shows the onset of inelastic
magnetic scattering above an energy transfer
of approximately 3 meV.
The low-energy onset of
the inelastic
magnetic scattering has been fitted with a Lorentzian convoluted with the
resolution function, yielding a value of
$$
E_{gap} = 3.7\pm 0.2 \  {\rm meV}
\eqno(3)
$$
for the gap.  This is
illustrated in Fig.2
by a dashed line in the upper graph and by a solid line in the
lower graph, which shows the data after subtraction of the elastic and
quasi-elastic components.  The smaller dashed line on the lower graph
represents the resolution function, and suggests
that the spectrum has an
intrinsic width, as expected if we are exciting a band rather than a
discrete level.

The open symbols on the upper
graph represent the data which was collected in the 2.5 m detectors at
$\phi = 130^o$.
At these high angles, and thus high Q values, there is no magnetic
scattering due to the fall-off of the ionic form factor, and the inelastic
scattering arises from single and multiple phonon scattering processes.
This data has been multiplied by an empirical scale factor of 0.2
(determined in previous experiments) to provide
an indication of the amount of non-magnetic background scattering underlying
the magnetic scattering data collected at low Q values.
The non-magnetic inelastic scattering contribution estimated in this way is
clearly very small, and there are no features coincident with the observed
magnetic scattering.  The non-magnetic background scattering estimated in this
manner has not been subtracted from the data presented in the lower plot.
This data shows a feature at 1.8 meV energy transfer which is also
visible in the scaled high angle data, confirming that the non-magnetic
background scattering is indeed as weak as estimated by the scaled high
angle data.

The resolution function of the spectrometer is made up of a contribution
from the moderator and a contribution from the transmission of the
chopper.  The former is described by a Gaussian convoluted with an
exponential, the later simply by a Gaussian.  The data below an energy
transfer of 4 meV has been fitted using the FRILLS least squares fitting
routine.\refto{rio} The elastic line
is well described by the resolution function, but there appears to be a
second broad component under the elastic line, which has been fitted with
a Lorentzian convoluted with the resolution function.  The origin of this
scattering is not yet understood, and may involve scattering from magnetic
defects in the sample.

No attempt was made to fit the scattering
intensity versus energy above
the gap using the ladder model, because there is an obvious discrepancy; the
observed scattering
is considerably stronger than expected
from numerical studies of $S(\vec Q,\omega)$ on small lattices,\refto{JRunpub}
which anticipate a rapid fall in intensity away from the gap.
This is the single feature of our data which does not appear to be in accord
with current expectations for the ladder model.

The overlap of the one- and two-magnon bands would give rise to
considerable structure in the scattering above the gap energy,
notably contributions
from two-magnon production at $\omega \approx 8$ and $17$ meV,
for which
we do see some evidence. This includes scattering
above 10 meV near
$Q=0.8$ \AA$^{-1}$ which is
evident in our 35 and 100 meV data.
The 100 meV data indicates
that this
higher energy, $Q\approx 0.8$ \AA$^{-1}$
scattering is strongest around 15 meV, and may therefore be due to the
two-magnon band expected at 17 meV. Due to the gaps in experimental coverage
we cannot be certain about the interpretation
of these higher-energy excitations.

If data were available across a broader $Q$ range,
and without the null region evident in Fig.1,
which is a consequence of the space
between the $2.5$ m and $4$ m banks of detectors, it should be possible to
extract the shape of the two branches and to study the
indications of anomalous behavior in $S(Q,\omega)$.  Such an experiment
is planned for the near future.
\vskip 0.4cm

\centerline{\bf III. Summary and Conclusions}
\vskip 0.25cm

In this note we present results from an inelastic neutron scattering
experiment on the antiferromagnet vanadyl pyrophosphate. We find evidence for
a band of triplet spin waves, with a minimum excitation energy of
$E_{gap}=3.7\pm 0.2$ meV at a momentum transfer near $Q=0.8$ \AA$^{-1}$.
These values are consistent with the expectations of the spin ladder model
of \vopon, which predicted $E_{gap}=3.9$ meV at $Q=0.813$ \AA$^{-1}$.
The dispersion relation $\omega(Q)$ of spin excitations from approximately
$0.6$ \AA$^{-1}$ to
$1.5$ \AA$^{-1}$ is evident in our data, and is also consistent with
expectations for the lowest one-magnon band in the ladder model.
These results confirm that \vopo to a good approximation is a
realization of the antiferromagnetic Heisenberg spin ladder.

There remain two outstanding problems with the interpretation of neutron
scattering from \vopon.
First,
the intensity of scattering above the gap
appears to be much stronger than expected from structure factor calculations
on small lattices.
Second, although the broad features of the excitation spectrum were observed in
this experiment, the details of the expected bands
above 10 meV are rather complicated,
and an experimental investigation will require
more complete coverage in $(Q,\omega)$.
Ideally this would involve
a scattering experiment using a single crystal.

\vskip 0.5cm
\centerline{\bf IV. Acknowledgements}
\vskip 0.25cm

We are grateful to J.Riera for communicating theoretical results
for the
magnetic structure factor $S(\vec Q ,\omega)$ of a spin ladder
prior to the experiment, and for
providing other unpublished results relating to
ladder and alternating chain models.
We also thank S.R.White for providing
unpublished results for the singlet-triplet gap in the ladder model,
M.E.Hagen for assistance with the data
visualisation,
J.Berlinsky, E.Dagotto, J.Greedan, H.A.Mook and H.Mutka
for related discussions, and M.A.Sokolov for a translation of Ref.[\cite{32}].
We acknowledge the provision of neutron beam time
by the EPSRC (formerly SERC).
This study was supported in part by the Quantum
Structure of Matter Project at Oak Ridge National Laboratory, funded by the
USDOE Office of Scientific Computation under contract DE-AC05-840R21400,
managed by Martin Marietta Energy Systems Inc.

\endpage

\references

\refis{Haldane} F.D.M.Haldane, Phys. Lett. 93A, 464 (1983).

\refis{Affleck}
I.Affleck, J. Phys. 1, 3047 (1989).

\refis{3}
E.Dagotto and A.Moreo,
Phys. Rev. B38, 5087 (1988); Phys. Rev. B44, 5396(E) (1991).

\refis{4}
S.P.Strong and A.J.Millis, Phys. Rev. Lett. 69, 2419 (1992).

\refis{5}
E.Dagotto, J.Riera and D.Scalapino, Phys. Rev. B45, 5744
(1992).

\refis{6}
T.Barnes, E.Dagotto, J.Riera and E.S.Swanson,
Phys. Rev. B47, 3196 (1993).

\refis{7}
J.Riera,
Phys. Rev. B49, 3629 (1994).

\refis{8} T.Barnes and J.Riera, {\it The Susceptibility and Excitation
Spectrum of (VO)$_2$P$_2$O$_7$ in Ladder and Dimer Chain Models},
ORNL / Rutherford Appleton Laboratory report ORNL-CCIP-94-04 / RAL-94-027
(1994), to appear in Phys. Rev. B.

\refis{9}
R.M.Noack, S.R.White and D.J.Scalapino,
{\it Correlations in a Two-Chain Hubbard Model}, University of California
(Irvine) report UCI-CMTHE-94-01 (1994).

\refis{10}
S.R.White, R.M.Noack and D.J.Scalapino,
{\it Resonating Valence Bond Theory of Coupled Heisenberg Chains},
University of California
(Irvine) report UCI-CMTHE-94-02 (1994).

\refis{11}
S.Gopalan, T.M.Rice and M.Sigrist, Phys. Rev. B49, 8901 (1994).

\refis{12}
M.Sigrist, T.M.Rice and F.C.Zhang,
{\it Superconductivity in a Quasi One Dimensional Spin Liquid}, (1994).

\refis{13}
H.Tsunetsugu, M.Troyer and T.M.Rice,
{\it Pairing and Excitation Spectrum in Doped $t-J$ Ladders},
Eidgen\"ossische Technische Hochschule report ETH-TH/94-01 (1994).

\refis{14}
D.Poilblanc, H.Tsunetsugu and T.M.Rice,
{\it Spin Gaps in Coupled $t-J$ Ladders},
Lab. de Physique Quantique report LPQTH-94-05 (1994).

\refis{15}
M.Azzouz, L.Chen
and S.Moukouri, {\it Calculation of the Singlet-Triplet Gap of the
Antiferromagnetic Heisenberg Model on the Ladder}, Universit\'e de Sherbrooke
report (1994).

\refis{16}
K.Hida, J. Phys. Soc. Jpn. 60, 1347 (1991).

\refis{17}
S.Takada and H.Watanabe,
J. Phys. Soc. Jpn. 61, 39 (1992).

\refis{18}
H.Watanabe, K.Nomura and S.Takada,
J. Phys. Soc. Jpn. 62, 2845 (1993).

\refis{19}
H.Watanabe,
{\it Numerical Diagonalization Study of an S=1/2 Ladder Model with Open
Boundary Conditions}, University of Tsukuba report.

\refis{20}
H.Watanabe,
{\it Haldane Gap and the Quantum Spin Ladder},
University of Tsukuba Ph.D. thesis (1994).

\refis{21}
J.C.Bonner, S.A.Friedberg, H.Kobayashi and B.E.Myers,
in {\it Proceedings of the 12th International Conference on Low Temperature
Physics}, ed.E.Kanda (Keigaku, Tokyo 1971), p.691.

\refis{22}
J.C.Bonner and H.W.J.Bl\"ote, Phys. Rev. B25, 6959 (1982).

\refis{23}
J.C.Bonner, S.A.Friedberg, H.Kobayashi, D.L.Meier and
H.W.J.Bl\"ote, Phys. Rev. B27, 248 (1983).

\refis{cuno1} K.M.Diederix, J.P.Groen, L.S.J.M.Henkens, T.O.Klassen
and N.J.Poulis, Physica 94B, 9 (1978).

\refis{cuno2} J.Eckert, D.E.Cox, G.Shirane, S.A.Friedberg
and H.Kobayashi, Phys. Rev. B20, 4596 (1979).

\refis{chc1}
D.B.Brown, J.A.Donner, J.W.Hall, S.R.Wilson, D.J.Hodgson and W.E.Hatfield,
Inorg. Chem. 18, 2635 (1979).

\refis{chc2}
W.E.Hatfield, J. Appl. Phys. 52, 1985 (1981).

\refis{SCO1} Z.Hiroi, M.Azuma, M.Takano and Y.Bando, J. Solid State Chem. 95,
230 (1991).

\refis{SCO2} M.Takano, Z.Hiroi, M.Azuma and Y.Takeda, Jpn. J. Appl. Phys. 7,
3 (1992).

\refis{LCO} R.J.Cava {\it et al.}, J. Solid State Chem. 94, 170 (1991).

\refis{26} E.Bordes, Thesis (Paris, 1973).

\refis{27} E.Bordes and P.Courtine, R\'eunions Annuelles Soc. Chim.
Fr. (Marseilles, 1973).

\refis{28} B.Kubias, Thesis
(Berlin, 1974).

\refis{29} H.Seeboth, G.Ladwig, B.Kubias, G.Wolf and B.Luecke,
Ukr. Chem. J. 43, 842 (1977).

\refis{30}
N.Middlemiss,
McMaster University Ph.D. thesis (1978).

\refis{31} E.Bordes and P.Courtine, J. Catal. 57, 236 (1979).

\refis{32}
Y.Gorbunova and S.A.Linde, Dokl. Akad. Nauk SSSR 245, 584 (1979).

\refis{33} D.C.Johnston and J.W.Johnson, J. Chem. Soc., Chem. Comm. 1720
(1985).

\refis{34} D.C.Johnston, J.W.Johnson, D.P.Goshorn and A.J.Jacobson,
Phys. Rev. B35, 219 (1987).

\refis{JRunpub} J.Riera, private communication.

\refis{SRWunpub} S.R.White, private communication.

\refis{MPMSCT} See for example
H.Mutka, C.Payen, P.Molinie, J.L.Soubeyeroux, P.Colombet and
A.D.Taylor, Phys. Rev. Lett. 67,  497 (1991), and references cited therein.

\refis{rio} R.Osborn, Rutherford Appleton Laboratory report RAL-91-011.

\endreferences

\vfill\eject
\noindent
Figure Captions
\vskip 0.25cm

\vskip 0.5cm
\noindent
Figure 1

\vskip 0.25cm
\noindent
Contour plot of the scattering intensity
from \vopo at 13 K using an incident energy of
25 meV.  Contours of intensity
run from 0.5 to 1.3 in steps of 0.1. (Arbitrary units.)
The one-magnon (solid line) and two-magnon (dashed line) bands predicted
by Barnes and Riera\refto{8} are superimposed on the data.

\vskip 0.5cm
\noindent
Figure 2

\vskip 0.25cm
\noindent
A constant $\phi$ scan across the predicted band minimum.  Data from eight
detectors which trace out the trajectories shown in the inset, have been
summed.
The upper graph shows the data (filled circles) and the fit as described in
the text, together with data from the high angle bank of detectors
(open circles), which have been multiplied by
an empirical scale factor of 0.2 to provide an estimate of
the non-magnetic background scattering.  The lower graph represents the
magnetic scattering after subtraction of the elastic and quasi-elastic
components.

\endpage
\vfill\eject\end